\documentclass[smallextended]{svjour3}       
\usepackage{graphicx}
\journalname{GERG}
\begin{document}

\title{Probing dark matter and dark energy through Gravitational Time Advancement.}



\author{Samrat Ghosh         \and
        Arunava Bhadra \and Amitabha Mukhopadhyay 
}


\institute{Samrat Ghosh \at
              High Energy $\&$ Cosmic Ray Research Centre, University of North Bengal, Siliguri, West Bengal, India 734013 \\
              \email{samrat.ghosh003@gmail.com}           
           \and
           Arunava Bhadra \at
             High Energy $\&$ Cosmic Ray Research Centre, University of North Bengal, Siliguri, West Bengal, India 734013 \\
						 \email{abhadra@nbu.ac.in}
						\and
						Amitabha Mukhopadhyay \at
             Department of Physics, University of North Bengal, Siliguri, West Bengal, India 734013 \\
						 \email{amitabha\_62@rediffmail.com}
}

\date{Received: date / Accepted: date}

\maketitle

\begin{abstract}
The expression of gravitational time advancement (negative time delay) for particles with non-zero mass in Schwarzschild geometry has been obtained. The influences of the gravitational field that describes the observed rotation curves of spiral galaxies and that of dark energy (in the form of cosmological constant) on time advancement of particles have also been studied. The present findings suggest that in presence of dark matter gravitational field the time advancement may take place irrespective of gravitational field of the observer, unlike the case of pure Schwarzschild geometry where gravitational time advancement takes place only when the observer is situated at stronger gravitational field compare to the gravitational field encountered by the particle during its journey. When applied to the well known case of SN 1987a, it is found that the net time delay of a photon/gravitational wave is much smaller than quoted in the literature. In the presence of dark matter field, the photon and neutrinos from SN 1987a should have been suffered gravitational time advancement rather than the delay. 

\end{abstract}

\section{Introduction}
\label{intro}
Light propagation in gravitational field leads to an extra time delay over the time required for light transmission between two points in Euclidean space, which is the well known gravitational or Shapiro time delay effect \cite{sha64},  \cite{sha66}. The observation of the time delay effect in the solar system constitutes one of the classical tests of general relativity. The difference in gravitational time delay between photon/gravitational waves and neutrinos or any other neutral particle with non-zero mass also has been used as a probe to examine the Principle of Equivalence \cite{lon88}and dark sector of the universe \cite{asa08}, \cite{sar16}. Presently the gravitational time delay effect is often employed to measure the masses of pulsars in binary systems \cite{dem10}, \cite{cor12}. 

Gravitational time delay is generally estimated by evaluating the additional coordinate time needed by a photon or a particle in a round trip journey in a gravitational field of a massive object over the coordinate time required in the absence of the gravitating object. However, the coordinate time difference is not a measurable quantity in a gravitational field; one needs to convert the coordinate time difference in to proper time difference which is a real measurable quantity. When such conversion is considered an opposite kind of effect, the so called gravitational time advancement (GTA) (negative time delay), is taken place if the observer is situated at stronger gravitational field in respect to the gravitational field encountered by the photon during its journey \cite{bha10}. The GTA effect is essentially caused by the fact that clock runs differently in gravitational field depending on the curvature. Note that when an observer is at weaker gravity field and is exploring time delay effect due to stronger gravity, such as time delay effect due to gravitational field of the Sun, the difference in coordinate time is roughly the same to the difference in proper time. That is why Shapiro effect is experimentally verified without any issue till now. 

The GTA of photons has been found to be affected by dark matter and dark energy \cite{gho15} and therefore, at least in principle, the measurements of GTA at large distances can verify the dark matter and a few dark energy models or put upper limit on the dark matter/energy parameters. The measurement of GTA also can be employed to discriminate the Gravity Rainbow (photons of different energies experience different gravity levels) from pure General Relativity \cite{den17}.

Like photons, particles having non-zero masses should also suffer GTA when the observer is at stronger gravitational field. Here we like to derive expression of GTA for particles with non-zero mass in Schwarzschild geometry. We further wish to examine the effect of the gravitational field that describes the observed rotation curve of spiral galaxies (in this paper we denote it as dark matter field) and the dark energy in the form of Cosmological constant on gravitational time advancement. The importance of the present investigation is many fold: It offers, at least in principle, to probe the presence of dark matter and dark energy, it constitutes a possible test of the GTA and it allows to estimate mass of a particle of unknown mass.   

The plan of the paper is the following. In the next section we shall present the basic formulation for calculating gravitational time advancement for a particle. In section 3 we shall estimate the GTA in a round trip journey by a particles under the influence of Schwarzschild geometry. In section 4 we shall study the effect of cosmological constant and dark matter gravitational field on GTA. We shall discuss the results in section 4 and conclude our findings in the same section.           

\section{Methodology}
\label{sec:1}
Consider the following scenario: An electromagnetic/gravitational wave or a particle is moving between the points A and B in a gravitational field due to a static spherically symmetric matter distribution as depicted in figure 1.  

\begin{figure}[t]
  \begin{center}
\includegraphics[width = 0.4\textwidth,height = 0.5\textwidth,angle=0]{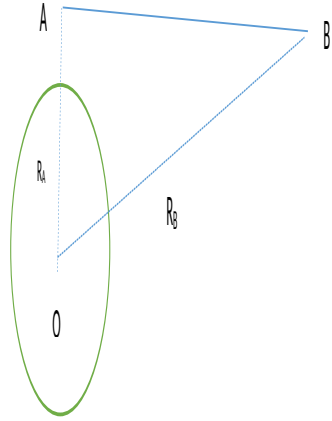}
\end{center}
\caption{Geometrical configuration of time delay/advancement of photon/particle in gravitational field. O is the Centre of the spherically symmetric mass distribution, A and B are two arbitrary points. $r_A$ and $r_B$ are radial distances of X and Y from O respectively.}
\label{Fig:1}
\end{figure}

We consider that the gravitational field is described by a general static spherically symmetric metric,

\begin{eqnarray}
ds^2=-\kappa(r) c^2 dt^2+\sigma(r)dr^2+r^2d\Omega^2  \;.
\end{eqnarray}
 
The geodesic equations for a test particle motion in equatorial plane under the influence of the space time given by Eq. (1) leads to the following relation 

\begin{equation}
\frac{\sigma(r)}{\kappa(r)^{2}} \left(\frac{dr}{dt}\right)^{2} + \frac{\alpha_1}{r^2} - \frac{c^2}{\kappa(r)} = -\alpha_2 c^2 \, , 
\end{equation}

where $\alpha_1$ ($\equiv \frac{r^4}{\kappa^{2}} \left(\frac{d\phi}{dt}\right)^2 $, and $\alpha_2$ ($\equiv \left(\kappa \frac{dt}{d\tau}\right)^{-2}$ for massive particle and $\alpha_2=0$ for mass less particle) are associated with the constants of motion, $\alpha_1$ is related to the angular momentum of the particle and $\alpha_2$ is related to the energy $\epsilon$ of the particle. At the distance of closest approach $r_o$, $\frac{dr}{dt}$ must vanish, which gives

\begin{eqnarray}
\alpha_1 = c^2 \left[- \alpha_2 + \frac{1}{\kappa(r_o)}\right]r_{o}^{2} \, ,
\end{eqnarray}

and $\alpha_2=\frac{m^2 c^4}{\epsilon^2}$, $m$ and and $\epsilon$ ($\equiv mc^2 \kappa \frac{d(t)}{d\tau} $) are the mass and energy of the particle. Hence the time required by a particle to traverse a distance from $r_o$ to $r$ is given by

\begin{eqnarray}
\Delta t \, (r,r_o)=\frac{1}{c} \int_{r_o}^{r}\sqrt{P(r,\alpha_2)} \, dr \, , 
\end{eqnarray}

where, 

\begin{eqnarray}
P(r,\alpha_2) = \frac{\sigma(r)/\kappa(r)}{\left[1-\alpha_2 \, 
\kappa(r) + \frac{{r_o}^{2}}{r^{2}}\left(\alpha_2 \,\kappa(r) 
- \frac{\kappa(r)}{\kappa(r_o)} \right ) \right]} \, .
\end{eqnarray}

Therefore the difference in proper time between transmission and reception in a round trip journey of the signal to be measured by the observer at $r_o$ is

\begin{eqnarray}
\Delta \tau = 2 \sqrt{\kappa(r_o)} \Delta t \, (r,r_o), 
\end{eqnarray}

Since the expression in Eq. (6) through Eq.(4) involves integration of the function $P(r,\alpha_2)$ which involves the metric coefficients $\sigma(r)$ and $\kappa(r)$, explicit expressions for $\sigma(r)$ and $\kappa(r)$ are required to proceed further. In the following sections we shall evaluate the proper time between transmission and reception for three different physically viable choices of $\sigma(r)$ and $\kappa(r)$. 

\section{GTA of a particle with non-zero mass in Schwarzschild geometry}
\label{sec:2}
In the Schwarzschild geometry i.e. when $\kappa(r) = \sigma(r)^{-1} = 1 - \frac{2\mu}{r}$ where $\mu = GM/c^2$, G is the gravitational constant and c is the speed of light, the coordinate time delay in round trip journey by a particle of mass m between A and B up to the first order accuracy of $\mu$ is given by

\begin{eqnarray}
\Delta t_m^{Sch} &=& \frac{2}{c \sqrt{1-\alpha_2}} [ \sqrt{r_A^2-r_o^2}+\sqrt{r_B^2-r_o^2}   \nonumber \\
&& + \frac{\mu\left(2-3 \alpha_2\right)}{\left(1-\alpha_2\right)} \ln\frac{\left(r_A+\sqrt{r_A^{2}-r_{o}^{2}}\right)\left(r_B+\sqrt{r_{B}^{2}-r_{o}^{2}}\right)}{r_{o}^{2}} \nonumber \\
&& + \frac{\mu}{\left(1-\alpha_2\right)}\left(\sqrt{\frac{r_{A}-r_{o}}{r_{A}+r_{o}}}+\sqrt{\frac{r_{B}-r_{o}}{r_{B}+r_{o}}}\right) ] \, ,
\end{eqnarray}
The first term in the right hand side of the above expression gives the special relativistic time take for the propagation whereas the rest of the terms are the \lq the Shapiro delay \rq in Schwarzschild spacetime. For a particle of mass m and energy $\epsilon$, $\alpha_2=\frac{m^2 c^4}{\epsilon^2}$. Hence the difference in proper time between transmission and reception of a particle of mass m from A to B and back to be measured by the observer at A reads

\begin{eqnarray}
\Delta \tau_{m}^{Sch} &=& \sqrt{B(r_A)} \Delta t_m^{Sch} \simeq \frac{2}{c \sqrt{1-\alpha_2}} [ \left(\sqrt{r_A^2-r_o^2}+\sqrt{r_B^2-r_o^2} \right) \left(1-\frac{\mu}{r_A} \right) +  \nonumber \\ 
&& \frac{\mu\left(2-3 \alpha_2\right)}{\left(1-\alpha_{2}\right)} \ln\frac{\left(r_{A}+\sqrt{r_{A}^{2}-r_{o}^{2}}\right)\left(r_{B}+\sqrt{r_{B}^{2}-r_{o}^{2}}\right)}{r_{o}^{2}} \nonumber \\ 
&& + \frac{\mu}{\left(1-\alpha_2\right)}\left(\sqrt{\frac{r_{A}-r_{o}}{r_{A}+r_{o}}}+\sqrt{\frac{r_{B}-r_{o}}{r_{B}+r_{o}}}\right)  ] \, .
\end{eqnarray}

Since both $\frac{\mu}{r_A}$ and Shapiro delay terms are small compare to special relativistic term, here we have ignored their higher order and cross terms. In the absence of the gravitating object (i.e. in flat space time) the time required by a particle of mass m and energy $\epsilon$ to travel between A and B is $\left(\sqrt{r_A^2-r_o^2}+\sqrt{r_B^2-r_o^2}\right)/\left(c\sqrt{1-m^2c^4/\epsilon^2}\right)$. Due to  gravitational effect this time is shorten by a factor $(1-\mu/r_{A})$ (first term in the right hand side of the above expression) which leads gravitational time advancement. The observed time will be smaller than the special relativistic time of propagation i.e. there will be a net GTA when the distance between A and B exceeds a certain value so that $\mu/r_A$ factor overcomes the Shapiro delay. The above expression thus gives the GTA (describes the situation when the gravitational time advancement effect overcompensates the Shapiro delay) for particles with mass $m$. The net GTA for massless particles such as photon can be readily obtained from the above equation by putting $\alpha_2=0$ (corresponding to $m=0$). 

If $r_B$ is much larger than $r_A$ and $r_{o}$, the expression for net GTA of particle with mass m can be approximated as 

\begin{equation}
\Delta \tau_{m}^{Sch} \approx \frac{2}{c \sqrt{1-\alpha_2}} r_{B} \left( 1  - \frac{\mu }{ r_A} \right) \, ,
\end{equation}

For relativistic particles ($\epsilon >> m$) and when $r_o \sim r_A$ the Eq. (9) reduces to

\begin{equation}
\Delta \tau_{m}^{Sch} \approx \frac{2 r_B}{c} \left[ \left(1-\frac{\mu}{r_A} \right)\left(1+\frac{m^2 c^4}{2\epsilon^2} \right) \right] \, .
\end{equation}
  
Therefore, the difference in arrival times after a round trip journey between particle with mass m and energy $\epsilon$ and photon those emitted at the same time reads

\begin{equation}
\Delta \tau_{m}^{Sch} - \Delta \tau_{\gamma}^{Sch} \approx \frac{m^2 c^3 r_B}{\epsilon^2} \left(1-\mu/r_A \right)  \, ,
\end{equation}

The first part in the right hand side of the above expression is the special relativistic effect whereas the second part is the GR correction.

Under the same conditions the difference in arrival times between particles with the same mass but different energies $\epsilon_1$ and $\epsilon_2$ with $\epsilon_2 > \epsilon_1$ is given by

\begin{equation}
\Delta \tau_{m}^{Sch}(\epsilon_2) - \Delta \tau_{m}^{Sch}(\epsilon_1) \approx m^2 c^3 r_B \left(\frac{1}{\epsilon_{2}^{2}} - \frac{1}{\epsilon_{1}^{2}} \right) \left(1-\mu/r_A \right)  \, ,
\end{equation}

Here an important point to be noted by examining the Eq. 7 that the sign of the expression of Shapiro time delay does not change for traveling from a stronger field to a weaker one and back again instead of traveling from a weak gravitational field to a stronger one and return back (the Shapiro delay is the same in both the situation). Rather a new effect, owing to the fact that that clock runs differently in gravitational field depending on the curvature, comes into play that leads to negative time delay or GTA in all the cases. The Shapiro delay mainly varies logarithmically with distance while the GTA varies linearly with distance. For a particle traveling from a weak gravitational field to a stronger one and return back magnitude of the negative time delay effect is much smaller than that of the Shapiro time delay, the resulting delay thus is a positive one. But when a particle travels from a stronger field to a weaker one and back again, the negative delay component starts dominating after a certain (small) distance, leading to a net GTA. 

We have not mentioned any particular particle so far, our results are very general, applicable to any particle with non-zero mass and even with zero mass. However, charged particles also suffer electromagnetic interaction and therefore, only stable neutral particles can be exploit to examine the GTA/Shapiro time delay effect in a realistic situation. Neutrons with life time around 15 minutes in its rest frame, can be utilized to test GTA/Shapiro delay in certain astrophysical situations not involving very large distances. Neutrinos are stable but their mass is not definitely known yet. Moreover the upper limit of their mass is too small so that the mass effect on GTA of neutrinos is very small. 
    
We would estimate the magnitude of the GTA effect for a simple situation as follows: Consider that photon and thermal neutron are simultaneously sent from the top of the Earth’s atmosphere towards the Moon where they (photon and neutron) are reflected back at the originating point. To survive without decay, the kinetic energy of the neutron has to be at least around 1 MeV. The Shapiro delay of photon and 1 MeV neutron in the mentioned case are $0.07$ ns and $0.07$ $\mu$s respectively whereas the GTA of photon and 1 MeV neutron will be $\sim 0.9$ ns and 1.8 $\mu$s respectively. The difference in arrival times between two neutrons, one with kinetic energy 1 MeV and other having kinetic energy 10 MeV will be about 1.6  $\mu$s.  The magnitude of the net GTA effect in the mentioned situation is thus well within the reach of the modern experiments.  

The future astrometric missions Beyond Einstein Advanced Coherent Optical Network (BEACON) \cite{tur09} or the GRACE Follow-On (GRACE-FO) \cite{pie08} are expected to detect the GTA effect employing laser beam from space craft. The mission BEACON will put six numbers of small spacecraft in a circular orbit of radius 80000 km and each spacecraft will be equipped with laser transceivers. Introduction of thermal neutron transceivers along with laser transceivers in such a future mission will lead to detect the GTA effect of particles.

\section{Effect of Dark sector on GTA of a relativistic particle}
\label{sec:3}
A wide variety astrophysical observations suggest that ordinary baryonic matter composes only $4.9\%$ of the matter in the Universe \cite{ade14}. The rest is mainly composed of dark energy ($68.3\%$) and dark matter ($26.8\%$) components of unknown nature \cite{ade14}. In this section we shall examine the effect of dark matter and dark energy on GTA. We shall consider the same physical scenario as depicted in figure 1. 

The presence of dark energy and dark matter lead to some modification of the Schwarzschild metric as the exterior space-time of a spherically symmetric mass distribution. Let us consider the following functional form of $\sigma(r)$ and $\kappa(r)$ in Eq. (1) 

\begin{eqnarray}
\kappa(r)=1-2\mu/r-\beta_1 r^n
\end{eqnarray}

and 

\begin{eqnarray}
\sigma(r)=1+2\mu/r+\beta_2 r^n
\end{eqnarray}

where n, $\beta_{1}$ and $\beta_{2}$ are constants. We shall consider the following cases:

case 1: The choice $n=1$, and $\beta_{1}=\beta_{2} = -\beta = -\left(5.42 \times 10^{-39} \frac{M_B}{M_{\odot}} + 3.06 \times 10^{-28}\right)$ ${\rm m}^{-1}$ (i.e. a linear potential), where $M_B$ is mass of baryonic matter in galaxy, has been found to describe well the observed flat rotation curves (with maximum extension upto extending around $100$ kpc) of a sample of 111 spiral galaxies  \cite{man11}, \cite{man12}. Since the radial extension of dark matter in a galaxy is not known, maximum radial distance of validity of the model can not be stated with certainty. But in general the model should not be extended to intergalactic scale.  

case 2: When $n=2$, $\beta_{1}=\beta_{2}=\Lambda/3$ the above metric represents the Schwarzschild-de Sitter (SDS) or Kotler space-time which is the exterior space time due to a static spherically symmetric mass distribution in presence of the cosmological constant $\Lambda$ with $\Lambda \sim10^{-52} {\rm {m}^{-2}}$ \cite{kot18}. 

The coordinate time required by a particle to traverse a round trip distance from $r_A$, which coincides with the distance of closest approach, to $r_{B}$ under the influence of space time geometry defined by Eqs. (1), (13) and (14 ) is given by \cite{sar16}

\begin{eqnarray}
\Delta t_{n} \left(r_{B}, r_A \right)  \approx \Delta t_m^{Sch} \left(r_{B},r_A \right) + \frac{1}{ \, c \, \sqrt{1-\alpha_{2}}} \left\{ 
      \left[\beta_1 + \beta_2 - \frac{\beta_1 \alpha_2}{\left(1-\alpha_2 \right)} \right] \, {\mathcal{I}}^1_n  
     - \frac{\beta_1 }{ \left(1-\alpha_2 \right)} \, {\mathcal{I}}^2_n   \right\} \, ,
\end{eqnarray} 

where, ${\mathcal{I}}^1_n$ and ${\mathcal{I}}^2_n$ are integrals defined by ${\mathcal{I}}^1_n = \int_{r_A}^{r_{B}} \frac{{r_{B}}^{n+1} \, dr}{\sqrt{\left({r_{B}}^2-r^2_A \right)}}$ and ${\mathcal{I}}^2_n = r_{A}^{2} \int_{r_A}^{r_{B}} \frac{r_{B} \left({r_{B}}^n - r^n_A \right) \, dr}{\left({r_{B}}^2-r^2_A \right) \sqrt{\left({r_{B}}^2-r^2_A \right)}}$. In the above equation $\alpha_2=\frac{m^2}{(1-2\mu/r_{o}-\beta_1 r^n)\epsilon^2}$ which is also to be used here in $\Delta t_m^{Sch}$. 

For $n=1$ and $n=2$ corresponding to DM and DE model respectively, we have analytical solutions of ${\mathcal{I}}^1_1$, ${\mathcal{I}}^2_1$ and ${\mathcal{I}}^1_2$, ${\mathcal{I}}^2_2$ which are given below

\begin{eqnarray}
{\mathcal{I}}^1_1 &=& \frac{r_B}{2} \sqrt{r_B^2-r_A^2} +  \frac{r_A^2}{2}\, \ln \frac{r_B+ \sqrt{r_B^2-r_A^2} }{r_A} , \nonumber \\
 && {\mathcal{I}}^2_1 = -r_A^2\sqrt{\frac{r_B-r_A}{r_B+r_A}} + r_A^2 \ln \frac{r_B+ \sqrt{r_B^2-r_A^2} }{r_A}  \, .
\end{eqnarray}
\begin{eqnarray}
{\mathcal{I}}^1_2 = \frac{1}{3} \sqrt{r_B^2-r_A^2} \left(r_B^2+2r_A^2 \right) \, , \nonumber \\
{\mathcal{I}}^2_2 = r_A^2 \sqrt{r_B^2-r_A^2} \, .
\label{12}
\end{eqnarray}

Thus for the dark matter model i.e. when $n=1$, $\beta_1=\beta_2=-\beta$ the proper time required for the travel by a particle with mass m and energy $\epsilon$ for the round trip travel between A to B as measured by an observer at A is given by

\begin{eqnarray}
\Delta{\tau}_m^{\beta} &\simeq& \Delta{\tau}_m^{Sch} - \frac{1}{ \, c \, \sqrt{1-\alpha_{2}}} [\left( \beta - \frac{\beta \alpha_2}{2(1-\alpha_2) } \right) \left(r_{B} \sqrt{r_B^2-r_A^2} +  r_A^2 \ln \frac{r_B+ \sqrt{r_B^2-r_A^2} }{r_A} \right) \nonumber \\    
 &&    - \frac{\beta r_A^2}{ \left(1-\alpha_2 \right)} \left( \sqrt{\frac{r_B-r_A}{r_B+r_A}} + \ln \frac{r_B+ \sqrt{r_B^2-r_A^2} }{r_A} \right) + \beta r_A \sqrt{r_B^2-r_A^2} ]
\end{eqnarray}

In the above expression we have ignored the cross terms between M and $\beta$ and higher order terms in $\beta$. It is noted from the above equation that $\beta$ reduces the net time advancement. 

In the presence of the cosmological constant ($n=2$, $\beta_1=\beta_2=\Lambda/3$), the proper time required for the travel by a particle with mass m and energy $\epsilon$ for the round trip journey between A to B as measured by an observer at A reads

\begin{eqnarray}
\Delta \tau_m^{\Lambda} &\simeq& \Delta \tau_m^{Sch} + \frac{1}{3 \, c \, \sqrt{1-\alpha_2}} [ 
      \left(2\Lambda - \frac{\Lambda \alpha_2}{1-\alpha_2}  \right)  \left(\frac{1}{3} \sqrt{r_B^2-r_A^2} \left(r_B^2+2r_A^2 \right) \right)  \nonumber \\
&&     - \frac{\Lambda }{ 1-\alpha_2 } \left( r_A^2 \sqrt{r_B^2-r_A^2} \right) -\Lambda r_A^2 \sqrt{r_B^2-r_A^2} ]
\end{eqnarray}		

When $r_B>>r_A$, considering only the leading order terms, for relativistic particles the Eqs.(18) and (19) respectively reduce to

\begin{eqnarray}
\Delta \tau_m^{\beta} \approx \frac{2 r_B}{c} \left[ \left(1-\frac{\mu}{r_A} - \beta r_B/2 \right)\left(1+\frac{m^2 c^4}{2\epsilon^2}(1-\beta r_A) \right) \right]\, .
\end{eqnarray}

and

\begin{equation}
\Delta \tau_m^{\Lambda} \approx \frac{2 r_B}{c} \left[ \left(1-\frac{\mu}{r_A} +\Lambda r_B^2 /9 \right)\left(1+\frac{m^2 c^4}{2\epsilon^2}(1+\Lambda r_A^2 /3) \right) \right]\, .
\end{equation}

The GTA of photons/GW can be obtained from the above expressions by putting $m=0$. Therefore, the difference in arrival times after one way journey (half of the round trip travel time) from B to A between particle with mass m and energy $\epsilon$ and photon/GW those emitted at the same time reads

\begin{equation}
\Delta \tau_m^{\beta} - \Delta \tau_{\gamma}^{\beta} \approx \frac{m^2 r_B c^3}{2  \epsilon^2} \left(1 -\beta r_B /2 \right)  \, ,
\end{equation}
 
\begin{equation}
\Delta \tau_m^{\Lambda} - \Delta \tau_{\gamma}^{\Lambda} \approx \frac{m^2 r_B c^3}{2  \epsilon^2} \left(1 + \Lambda r_B^2/9 \right)  \, ,
\end{equation}

In the expressions for GTA of particles the first order effects of flat rotation curve and cosmological constant appear separately from the contribution of mass (Schwarzschild term) as revealed from Eqs. (18) to (21). Since the contribution of dark matter and dark energy are visible only at large distance scales, neutrons are not suitable for probing the dark matter/energy through GTA effect of particles. Neutrinos seem the only option in this regards.  

Another pertinent issue is that getting reflecting back a particle at the Earth from a large distance away is not a realistic idea. So instead of two way motion, we need to consider just one way motion. Measurement of GTA through one way motion can be performed, at least in principle, by sending light/particle from artificial satellite/space station to the Earth. Since the time of emission from a distant source is not known, measurement of GTA or Shapiro delay from one way travel is not possible in such cases. Instead the measurement of difference of arrival times of two particles (or a particle and a photon or two same kind of particles but with different energies) gives an opportunity to test GR and dark matter/energy models provided the relative time of emission of the particles is known within a small uncertainty.       

In the next section we shall see how the GTA effect alters the prevailing result of Shapiro time delay of the neutrinos from SN-1987. We shall also estimate the magnitude of dark matter contribution on the GTA of neutrinos from SN 1987.

\section{Discussion and conclusion:}
\label{sec:4} 
In Schwarzschild space time particles with non-zero mass suffers GTA when the observer is at stronger gravitational potential compare to the gravitational field encounter by the particle during its journey. The net GTA of particles with non-zero mass is found smaller than that of photons/GW. Due to lower speed, particles with non-zero mass should arrive later than the photon/GW if both were departed at the same instant from the source and the delay of particles with respect to photons can easily be estimated using special relativity. The gravitational time delay enhances the delay for particles with non-zero mass. The net delay in arrival time of relativistic particles, however, reduces to half of the gravitational time delay when proper time of the observer is taken into account.        

The dark matter field under the framework of conformal gravity leads to larger GTA. More importantly the GTA is influenced by the dark matter gravitational field at the source position. Thus if the source is located at large distance away (at the outskirt of the galaxy), the dark matter contribution to GTA can be quite large. Interestingly in the presence of dark matter field the prevailing condition for GTA that the observer has to be in stronger gravitational field is no more required. In the dark matter field the net GTA of particles with non-zero mass is found larger than that of photons/GW. 

In contrast to dark matter field effect the cosmological constant (dark energy) is found to reduce the magnitude of GTA which could be due to the repulsive nature of cosmological constant. Similar to dark matter case the contribution of cosmological constant to time delay can be large because the gravitational field due to cosmological constant at the source position contributes in the net delay.     

When the distance of the source is quite large compare to the observer distance from the gravitational object the GTA for particles with non zero mass is proportional to square of particle mass and  goes inversely with the square of the energy of the particles. So measurement of GTA can be exploit to evaluate mass or put limit on the mass of particles with unknown mass, at least in principle. Another relevant issue is that how far the dark matter halo extends to? The stability criterion can severely constrain the extent of the H1 gas in a galaxy and thereby leads to some testable upper limit on the size of a galaxy \cite{nan12}. The GTA effect can in principle be exploit to probe the extension of our galaxy.  

To exemplify the points stated above we consider the case of photons and neutrinos from the well known supernovae 1987A in the Large Magellanic Cloud. The neutrinos from SN `1987A arrived about four hours earlier than the appearance of the optical counterpart. Since the observer at the Earth is at higher gravitational field of the galaxy for the propagation of photons and neutrinos from the supernovae 1987A to the Earth, one needs to consider the proper time for evaluating the true time delay.  

The SN1987A is located at a distance about 50 Kpc \cite{pan98} and the travel time of a photon from SN1987a to the Earth is about $1.62 \times 10^5$ years. Considering that the total mass of the galaxy inside 60 kpc is about $6 \times 10^{11} \; M_{\odot}$ and the distance between the Earth and Center of the galaxy is about 12 kpc, the gravitational time delay (without considering proper time) experienced a photon while traveling from SN1987a to the Earth is about $1.2 \times 10^7$ seconds \cite{lon88}, \cite{bos88}. After considering the proper time interval and treating the galactic gravitational field as purely Schwarzschild in nature, the net delay will be nearly $2.85 \times 10^6$ seconds (here $r_B$ is not much larger than $r_o$ and hence the full expression as given in Eq. (9) needs to apply). So there is no time advancement in this case but the net gravitational delay is nearly an order less than that reported earlier \cite{lon88}, \cite{bos88}. If we consider the dark matter model described by case 1 of Eqs. (13) and (14), and assuming baryonic mass of the galaxy is about $16 \%$ of the total galactic mass the net delay for a photon will be $- 6.2 \times 10^6$ seconds i.e. there will be nearly half an year time advancement instead of time delay. At the distance of SN1987a, the effect of cosmological constant is quite small and its contribution ($\sim 240 \; s$) to the net gravitational time delay thus can be ignored. 

If we turn to SN1987a neutrinos, a major issue is that despite a huge progress in neutrino physics over the last three decades or so, the definite mass of the three neutrinos: electron, muon and tau neutrinos (and antineutrinos) are still unknown though experimental evidence of neutrino oscillations suggest that they are not massless. The cosmological observations give an upper bound on the sum of the active neutrinos $\sum m_\nu^{i} < 0.23 $ eV, \cite{ade14} here the superscript i denotes the mass eigenstate of neurinos. The Lyman alpha forest power spectrum suggests more stringent limits $\sum m_\nu^{i} < 0.12$ eV \cite{pal15}. The energy of the detected neutrinos from SN1987a is of the order of 10 MeV. Therefore, there will be no significant difference in time of arrival between photon and neutrinos emitted at same point of time, the correction term due to mass is less than a nano-second; much less than the intrinsic error . 

In the above analysis we assumed that metric parameters are identical for all the particles following the Einstein equivalence principle.  To examine a possible violation of Einstein equivalence principle one usually employ the post-parameterized Newtonian (PPN) metric i.e. $\kappa(r) = 1 - \frac{2\mu}{r}$ and $\sigma(r) = 1 + \frac{2 \gamma_i \mu}{r}$, (up to the accuracy of $\mu$) where $\gamma_i$ is the first PPN parameter that can be different for different particles, the subscript i denotes species of the particle. $\gamma$ is unity in general relativity, zero in the Newtonian theory. The observations suggests $\gamma$ is very close to 1 
\cite{ber03}. 
For the PPN metric the difference in proper time between transmission and reception of a particle of mass m from A to B and back to be measured by the observer at A reads

\begin{eqnarray}
\Delta \tau_{m}^{PPN} \simeq \frac{2}{c \sqrt{1-\alpha_2}} [ \left(\sqrt{r_A^2-r_o^2}+\sqrt{r_B^2-r_o^2} \right) \left(1-\frac{\mu}{r_{A}} \right) +  \nonumber \\ 
 \frac{\mu\left(1+\gamma_i-(2 +\gamma_i)\alpha_2\right)}{\left(1-\alpha_{2}\right)} \ln\frac{\left(r_{A}+\sqrt{r_{A}^{2}-r_{o}^{2}}\right)\left(r_{B}+\sqrt{r_{B}^{2}-r_{o}^{2}}\right)}{r_{o}^{2}} \nonumber \\
+ \frac{\mu}{\left(1-\alpha_2\right)}\left(\sqrt{\frac{r_{A}-r_{o}}{r_{A}+r_{o}}}+\sqrt{\frac{r_{B}-r_{o}}{r_{B}+r_{o}}}\right)  ] \, .
\end{eqnarray}     

and therefore, when $r_A \simeq r_o$ the difference in arrival times after a round trip journey between a relativistic particle with mass m and energy $\epsilon$ and a photon those emitted at the same time reads

\begin{eqnarray}
\Delta \tau_{m}^{PPN} - \Delta \tau_{\gamma}^{PPN} \simeq \frac{2}{c} [\sqrt{r_B^2-r_A^2} \left(1-\frac{\mu}{r_{A}} \right)\frac{m^2 c^4}{2\epsilon^2} + \mu ln \frac{r_B+\sqrt{r_B^2-r_A^2}} {r_A} \left(\gamma_{m}  
- \gamma_{\gamma} \right) \nonumber \\ + 3\mu \sqrt{\frac{r_B-r_A}{r_B+r_A}} \frac{m^2 c^4}{2\epsilon^2} ]\, ,
\end{eqnarray} 

Since neutrino mass is very small, the middle term of the right hand side of the above equation will dominate and hence effectively one gets the same expression that was used in \cite{lon88} to examine the Einstein equivalence principle using SN 1987A data considering neutrinos are massless particle.   

The recent detection of a few gravitational wave transients from sources at large distances creates better opportunity to examine the gravitational time advancement and its consequences. The gravitational waves and neutrinos are expected to emit within a short period (few seconds at most) of time from such binary black hole/neutron star coalescence or from supernova explosions. The observation of arrival time difference between gravitational wave and neutrinos from such large distance sources may provide an independent way to constrain on the mass of the neutrinos.

In conclusion in the present work we have obtained expressions for GTA of particles in Schwarzschild geometry for the first time by considering proper time interval of propagation of a particle with non-zero mass between two points in a gravitational field. Out findings suggest that the gravitational time advancement may take place when the observer is situated at stronger gravitational field compare to the gravitational field encountered by the particle during its journey. Subsequently we study the effect of dark matter and dark energy on gravitational time advancement. It is found that dark matter leads to larger gravitational time advancement whereas dark energy always produces time delay.  We have demonstrated how the present findings can be tested in a real observational situation. Finally after applying our findings to neutrinos (and photons) from SN 1987, we have shown that the net time delay of a photon/gravitational wave is much smaller than quoted in the prevailing literature due to GTA effect. 

Recently ICECUBE experiment and Fermi telescope detected neutrinos and photons within a short time period from BLAZER TXS 0506+056 \cite{ice18}, \cite{ice18a}. More such kind of detection from various sources are expected in near feature. The present findings will have direct application to test various underlying physics related issues of GR and particle physics from the measurement of the difference in time of arrivals of photons/gravitational wave and neutrinos from such astrophysical sources.  

%
%

\begin{acknowledgements} 
The authors would like to thank an anonymous reviewer for insightful comments and very useful suggestions that helped us to improve and correct the manuscript.
\end{acknowledgements}



\end{document}